\documentclass[11pt]{article}
\usepackage{times}
\usepackage{geometry}
\usepackage[utf8]{inputenc}
\usepackage{enumitem,amssymb}
\usepackage{ragged2e}
\newlist{thematic}{itemize}{8}
\setlist[thematic]{label=$\square$}
\usepackage{pifont}
\usepackage{fancyhdr}
\usepackage[USenglish]{babel}
\usepackage[utf8]{inputenc}
\usepackage[T1]{fontenc}
\usepackage{epsfig}
\usepackage{wrapfig}
\usepackage{color}

\usepackage[vflt]{floatflt}
\usepackage{longtable}
\usepackage{caption}
\usepackage{jheppub}   
\usepackage[dvipsnames,svgnames,x11names]{xcolor}
\usepackage[all]{hypcap} %Links go to figures; breaks on deluxetables (use \capstartfalse \capstarttrue to fix it)
\usepackage{amssymb,amsmath}
\usepackage{longtable}
\usepackage{amssymb}
\usepackage{amsmath, xspace}
\usepackage{graphics,graphicx} 
\usepackage{rotating}
\usepackage{color}
\usepackage{xcolor}
\usepackage{multirow}
\usepackage{subfigure}
\usepackage{sectsty}
\usepackage[outercaption]{sidecap}
\sidecaptionvpos{figure}{c}

\usepackage{enumitem}
\setlist[enumerate]{itemsep=0pt, parsep=0pt}
\setlist[itemize]{itemsep=0pt, parsep=0pt}

\definecolor{DarkGreen}{rgb}{0.0, 0.3, 0.0}
\definecolor{purple}{rgb}{0.5, 0.0, 0.5}
\definecolor{red}{rgb}{1, 0.0, 0.0}
\definecolor{green}{rgb}{0, 1.0, 0.0}

\def\3he{$^3{\rm He}$}

\def\lsim{\mathrel{\lower2.5pt\vbox{\lineskip=0pt\baselineskip=0pt
           \hbox{$<$}\hbox{$\sim$}}}}

\def\gsim{\mathrel{\lower2.5pt\vbox{\lineskip=0pt\baselineskip=0pt
           \hbox{$>$}\hbox{$\sim$}}}}

\begin{document}
\raggedright
\huge
The Emergence of Prebiotic Chemistry in the ISM  
\linebreak
\bigskip
\normalsize
\justifying

%\cc{\begin{itemize}
%    \item ESO's guidelines and Submission details website: \href{https://next.eso.org/call-for-white-papers/}{https://next.eso.org/call-for-white-papers/}
%    \item Submission Deadline: 15 Dec 2025
%    \item ESO will not make the white papers public but the authors can post them on arXiv.
%\end{itemize}}

%\raggedright
%\Large
%\cc{Notes on this AtLAST white paper template} 
%\linebreak
%\normalsize

%\cc{As stated on the ESO Expanding Horizons Call for white papers website, they are soliciting white papers from the community (at all career stages) to encourage broad discussions across the community and help identify future challenges.}
%\cc{To help coordinate our efforts, the AtLAST team has put together this latex template for internal use, and have decided to make it available more generally for anyone who may find it useful. }
%\cc{Some of the blue text guidance in each of the suggested sections is there to help some of our more junior team members understand what to expect to need to include (and at what level). }
%\cc{Any suggestions made here reflect how we are intending to answer the call issued by ESO.}
%\bigskip

\textbf{Authors:} 
%\cc{Note: first 3 authors must be from ESO countries}
Izaskun Jim\'enez-Serra (ijimenez@cab.inta-csic.es, Centro de Astrobiolog\'{\i}a (CAB, Spain));
Giuliana Cosentino (Institute de Radioastronomie Millim\'etrique, IRAM, France); Francisco Montenegro-Montes (Universidad Complutense de Madrid, UCM, Spain); Laura Colzi (CAB);  V\'{\i}ctor M. Rivilla (CAB); Miguel Sanz-Novo (CAB); Marta Rey-Montejo (CAB); David San Andr\'es (CAB); Sergio Mart\'{\i}n (European Southern Observatory, ESO, Chile); Shaoshan Zeng (RIKEN, Japan); Am\'elie Godard Palluet (CAB); Miguel A. Requena-Torres (Towson University, USA); Germ\'an Molpeceres (Instituto de F\'{\i}sica Fundamental - IFF, Spain); Pamela Klassen (UKRI STFC, UK); Doug Johnston (NRC-Herzberg Institute, Canada); Francesco Fontani (Osservatorio di Arcetri, Italy); Silvia Spezzano (Max Plank Institute for Extraterrestrial Physics - MPE, Germany); Elena Redaelli (ESO, Germany); Juris Kalvans (Venstpils University, Latvia); Yuri Aikawa (University of Tokyo, Japan); Bel\'en Tercero (Observatorio de Yebes - OAN, Spain); Pablo de Vicente (Observatorio de Yebes - OAN, Spain); Serena Viti (University of Leiden, The Netherlands); Emilio J. Cocinero (UPV/Biofisika Institute, Spain); Aran Insausti (UPV/Biofisika Institute, Spain)
\linebreak

\textbf{Science Keywords:} 
%\cc{Note: THIS IS NOT A REQUIREMENT FROM ESO. We could list here a few (3-6) of the keywords ESO uses in its proposal system here to help the Senior Science Committee understand what to expect from the white paper. An overview of ESOs proposal keywords used in the recent P117 call can be found in \href{https://www.eso.org/sci/observing/phase1/p117/CfP117.pdf}{Appendix A of this document}.}
ISM: clouds; ISM: molecules; Galaxy: center; stars: formation
\linebreak

 \captionsetup{labelformat=empty}
\begin{figure}[h]
   \centering
\includegraphics[width=.7\textwidth]{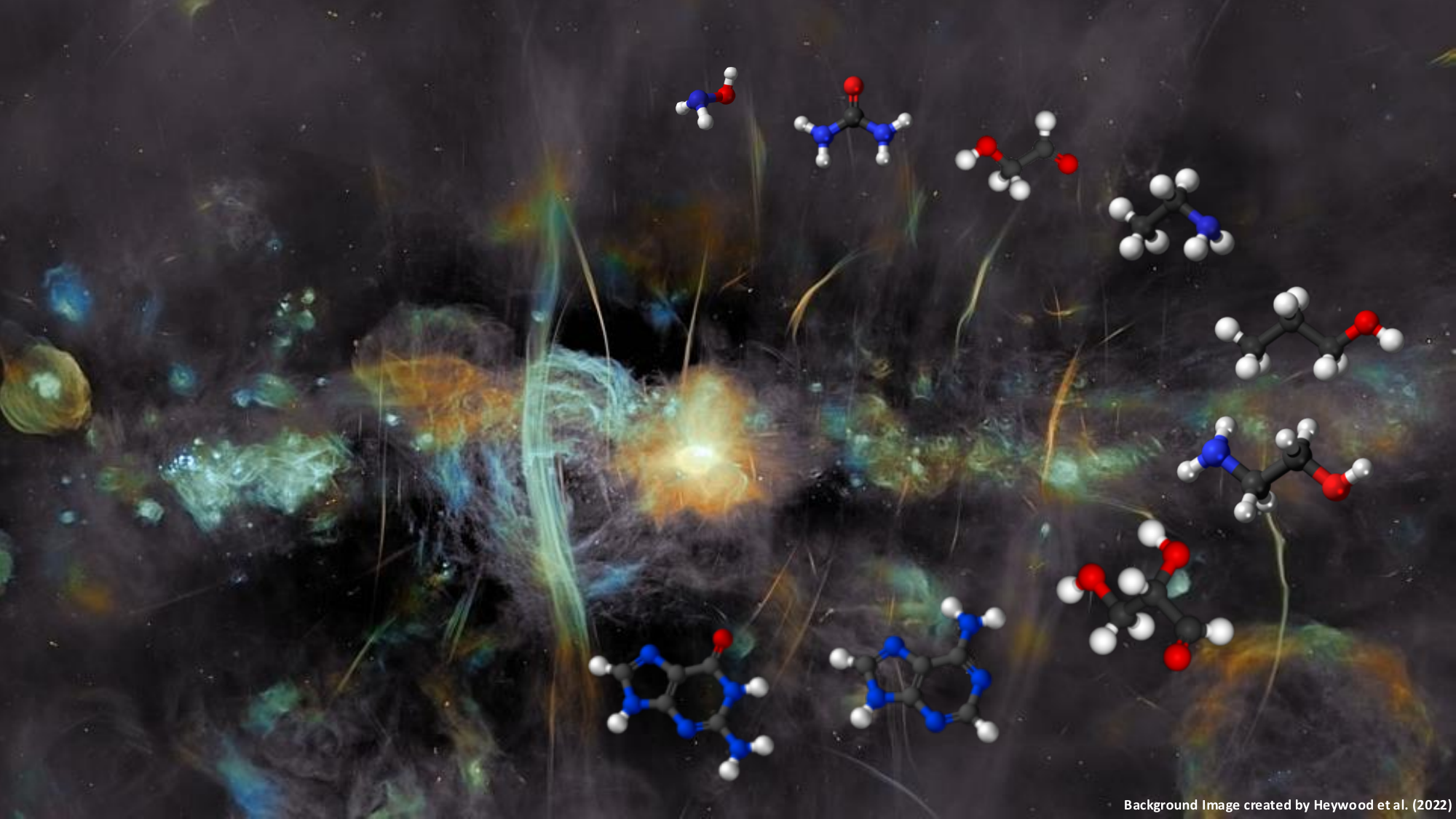}
   \caption{}
\end{figure}
\vspace{-15mm}

% this will reset the figure counter, so the text has figure 1.
\setcounter{figure}{0}
\captionsetup{labelformat=default}

%\justify
%\textbf{Abstract:} 

\pagebreak

%\cc{Below are some suggested section headings for the main 3-pages long document. Modify/delete/disregard as needed to best suit your white paper.} \cc{A reminder that the request from ESO is that the white papers should include:}
%\cc{\begin{itemize}
%    \item a focus on one (or group of) science question(s) and explain why this needs a facility we do not expect to have by the 2030s
%    \item a short ($<$ half page) description of what technology developments / data handling requirements may be needed.  %not looking for a detailed description of a facility
%\end{itemize}}

\section*{Abstract}
%\cc{Suggest to keep it short, 4-5 lines}
%AtLAST's superb sensitivity, multi-band observations and large FoV will enable the search for the building blocks of life in molecular clouds across multiple galactic environments and in external galaxies. This science case will allow us to establish whether prebiotic chemistry is a natural outcome of interstellar chemistry, representing a viable source of prebiotic feedstock essential for the emergence of life on early Earth and elsewhere in the Universe.
Contrary to popular belief, the interstellar medium (ISM) is not empty; it is filled with atoms, dust particles, and molecules. Some of these molecules may have been the very “building blocks of life” that, delivered to Earth via comets and meteorites, could have given rise to Life itself. A large-area single-dish telescope with superb sensitivity, field-of-view and multi-band instruments will allow us to explore the limits of chemical complexity in the interstellar medium, across our Galaxy and in external galaxies, determining whether amino acids, sugars, or RNA/DNA nucleobases can form in space.

\section{Scientific context and motivation}
%\cc{Target audience: expert astronomers, but not necessarily in your field. Could be good to keep the motivation to a similar level as the background section of an ALMA large proposal.}
The origin of life is one of the great unsolved problems in the field of the natural sciences. Our current understanding is that life emerged in an environment with an atmosphere, water, and continental masses where sufficient nutrients were available for the formation and subsequent evolution of the first proto-biological systems. These proto-biological nutrients (mainly biopolymers such as RNA, DNA, proteins, sugars, and phospholipids) are commonly regarded as “building blocks of life”. It is believed that they were synthesized abiotically from simpler prebiotic compounds\footnote{Prebiotic molecules are those involved in theories of the origin of life (see e.g. Kitadai \& Maruyama 2018).} present in aqueous environments on early Earth. However, the origin of those simpler organic compounds remains uncertain. 

The majority of those nutrients formed endogenously on the surface of a primitive Earth. However, it is currently believed that a significant fraction of those compounds might have formed exogenously in the proto-solar nebula, and were subsequently transferred to planetesimals and cometesimals (e.g. Ohishi 2019). The impact of these bodies on the surface of early Earth during the Late Heavy Bombardment period (between 4.1 and 3.8 billion years ago) made those prebiotic organic compounds available for the first biochemical reactions, playing a decisive role in the origin of life. This hypothesis is supported by the recent discovery within asteroids Ryugu and Bennu of all five nucleobases of DNA and RNA, and of 14 out of the 20 proteinogenic amino acids (Oba et al. 2023; Glavin et al. 2024). Nitrogen-15 isotopic enrichments indicate that these organic compounds were formed before the Solar-System in a cold molecular cloud. But, {\it what compounds were formed within the initial proto-solar nebula? And how complex can interstellar chemistry become? Does this chemical complexity permeates the universe?}

%In the following, we explain recent advances in our understanding of the emergence of chemical complexity in the interstellar medium (ISM) and present how AtLAST will be a key instrument in the quest of the building blocks of life in space.

\vspace{-3mm}
\section{Science case}

%\cc{Suggestions for content:}
%\cc{\begin{itemize}
%    \item Focus on one (or a group of) science question(s) of interest
%    \item 1-2 catchy self-explanatory Figures/graphics with captions to help explain your points.
%    \item Explain why this science case cannot be done without AtLAST (why it cannot be done with any other current or planned facility?)
%\end{itemize}}

\begin{figure*}
\begin{center}  \includegraphics[width=0.8\linewidth]{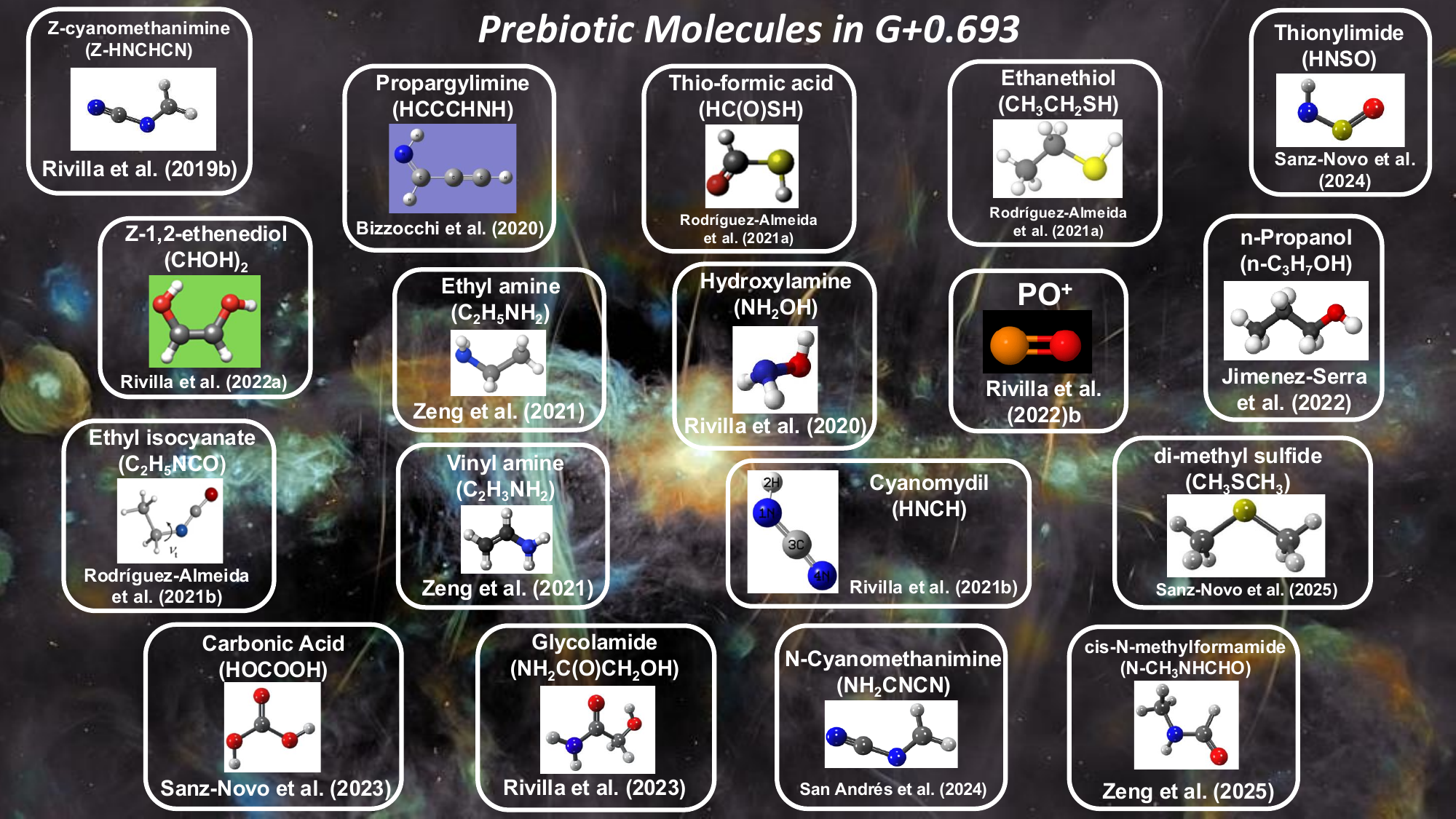}
\caption{Sample of prebiotic molecules detected toward the Galactic Center cloud G+0.693 including precursors of ribonucleotides, amino acids, sugars, proto-proteins and proto-lipids (see recent review by Jim\'enez-Serra 2025). Background image: SARAO, Heywood et al. (2022) / J. C. Mu\~{n}oz-Mateos.}
\label{fig:prebiotic}
\end{center}
\end{figure*}

The advent of high-sensitivity instrumentation at the IRAM 30m and Yebes 40m telescopes has enabled ultrasensitive and broadband spectral surveys that have fully sampled the atmospheric windows at 7mm, 3mm, 2mm and 1mm. These surveys have revealed 
a wide variety of prebiotic molecules in the ISM, which include precursors of ribonucleotides (the smallest units of RNA), sugars, amino acids, proto-proteins and proto-lipids (see Jim\'enez-Serra et al. 2025, for a recent review). 
Figure \ref{fig:prebiotic} shows a sample of the prebiotic compounds recently discovered in the ISM. Some examples are: hydroxylamine (NH$_2$OH), a precursor of ribonucleotides (Rivilla et al. 2020); ethanolamine (NH$_2$CH$_2$CH$_2$OH), the simplest phospholipid head group in cell membranes (Rivilla et al. 2021); n-propanol (n-C$_3$H$_7$OH), a precursor of fatty alcohols (Jimenez-Serra et al. 2022); and carbonic acid (HOCOOH), precursor of amino acids and lipids (Sanz-Novo et al. 2023). 

These detections have been achieved toward the molecular cloud G+0.693-0.027, located in the Galactic Center (hereafter G+0.693). This source is an excellent laboratory for the discovery of new interstellar species (and in particular, of complex organic molecules or COMs) because it is one of the richest chemical repositories in our Galaxy. Its molecular emission is sub-thermally excited (excitation temperatures of T$_{\rm ex}$=7-15 K versus gas kinetic temperatures of T$_{\rm kin}$=70-150 K; Zeng et al. 2018), leading to lower-frequency transitions from these COMs since only low molecular energy levels are populated. This significantly reduces the levels of line blending and line confusion in the observed spectra as the lower frequencies are less crowded by simpler species. In addition, the low-energy transitions of prebiotic molecules are brighter at lower frequencies because only their lowest energy levels can be populated at the low T$_{\rm ex}$ measured in G+0.693. The discovery of all these complex organics of prebiotic interest in this Galactic Center molecular cloud suggests that the ISM is a viable source of prebiotic compounds that could have contributed to the first steps in the complex process of the origin of life. This field is just starting up and we are witnessing the {\it tip-of-the-iceberg}. %Therefore, {\it we currently ignore what the limits of chemical complexity are, and whether the actual building blocks of life such as sugars, amino acids and even nucleobases, are present in the ISM.} 

\begin{figure*}
\begin{center}  \includegraphics[width=1.0\linewidth]{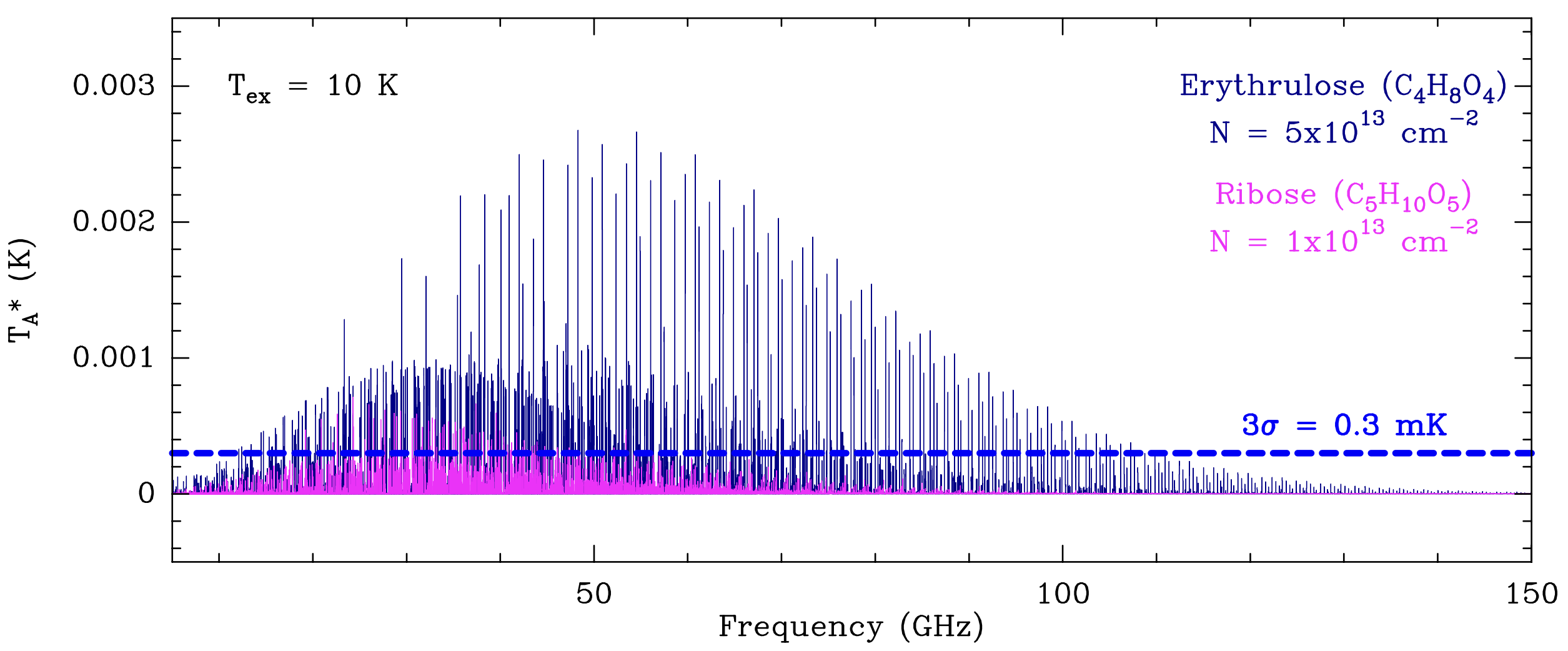}
\caption{Predicted spectra of erythrulose and ribose for the molecular excitation conditions measured in G+0.693. The blue dashed line indicates the 3$\sigma$ level = 0.3 mK needed for this science.}
\label{fig:ribose}
\end{center}
\end{figure*}

The spectral surveys already carried out with the IRAM 30m and Yebes 40m telescopes have devoted hundreds of hours to achieve average rms noise levels $\sim$1 mK. However, as shown in Figure$\,$\ref{fig:ribose}, the detection of sugars with four and five carbon atoms (C4 and C5 sugars) requires a sensitivity $\geq$10 times higher (i.e. rms noise levels $\sim$0.1 mK). %This rms noise level would require a prohibitive amount of integration time at current single-dish facilities ($\sim$0.8-1.1 yr at the Yebes 40m and IRAM 30m telescopes). 
%However, AtLAST can achieve that sensitivity in just 94 hours of integration time ($\sim$0.2 mJy for a velocity resolution of 1.5 km s$^{-1}$).
A large-area single-dish telescope with superb sensitivity is therefore needed. Simultaneous multi-frequency observations at multiple atmospheric windows (at 7mm, 3mm, 2mm and 1mm) would be a plus since it would provide information about hundreds of rotational transitions for every molecular species (COMs and simpler ones), essential to robustly identify prebiotic COMs and to determine their abundance. This is a unique capability of AtLAST. 

This sensitivity will not only enable searches of biomolecules in space in the Galactic Center but also across the Galaxy. The large-scale spectral line mapping of the Galactic disk with AtLAST (see White Paper by Klaassen et al. and Klaassen et al. 2024) will find new targets for complementary deep spectral surveys. Of particular interest are regions in the Outer Galaxy characterized by lower metallicity conditions, where COMs have been detected (Shimonishi et al. 2021; Fontani et al. 2024). High-sensitivity observations carried out toward the Magellanic clouds (LMC and SMC; metallicities of $\sim$0.5 and $\sim$0.25, respectively) have also revealed the presence of COMs in these galaxies, indicating that COM production is independent of metallicity (Shimonishi 2024). 
This opens the possibility of studying the presence of prebiotic COMs not only in the outer Galaxy and Magellanic clouds but also in external galaxies, which will allow us to {\it establish whether prebiotic chemistry is a natural outcome of interstellar chemistry.}   

%\vspace{-0.5cm}
\section{Technical requirements}
For this science case, we require high-sensitivity, broadband observations carried out at frequencies mainly between 30 and 50 GHz and covering a large FoV. AtLAST sensitivity will be superb at these frequencies, which will open a new avenue for the discovery of new prebiotic species. The requested sensitivity of 0.1 mK can be achieved by AtLAST in less than 100 hours, clearly surpassing the capabilities of current telescopes that require over a year to achieve a similar sensitivity.   
%in space and for our understanding on the processes involved in the emergence of chemical complexity and in its inheritance into planetary systems. 
%The primary target of this science case are molecular clouds. 
The emission of COMs in Galactic Center clouds is extended across several parsecs (see e.g. Li et al. 2017, 2020) and hence, single-dish telescopes are the best instruments to observe this type of emission since interferometers filter significant amounts of flux. For a distance of 8 kpc (Galactic Center), a beam resolution of $\sim$38-27$"$ at 35-50 GHz will be sufficient to probe detailed information on the morphology and kinematics of the molecular emission at parsec/sub-parsec spatial scales across a FoV of $\sim$2 deg covered by a heterodyne "multi-beam array". If the 10-mm array is "sparse" (filling all/part of the FoV with "holes" among receivers) this could be planned as just "staring" or "wobbler" observations (see AtLAST Memo \#2), where some of the beams in the FoV would target the regions of interest and some others would target OFF (empty) positions. If the multi-beam array filled tightly (e.g. with separations 2F$\lambda$) some area of the focal plane, one would need fast OTF- or daisy-like pattern observations to map the area of interest.
The typical linewidths in Galactic Center molecular clouds are of a few tens of km s$^{-1}$. Velocity resolutions of $\sim$1-2 km s$^{-1}$ are sufficient for our goals, although note that the molecular line profiles observed in Outer Galaxy clouds (with similar T$_{ex}$ as in Galactic Center clouds) can be as narrow as 1-2 km s$^{-1}$, requiring higher velocity resolutions ($\sim$0.2 km s$^{-1}$). We stress that the frequency range 30-50 GHz represents the 'sweet spot' for detecting prebiotic molecules (see Figure \ref{fig:ribose} since their spectra peak at these frequencies given their sub-thermal excitation in molecular clouds. Therefore, AtLAST is the only instrument that can carry out this science because these frequencies will not be covered by other facilities such as the Large Millimeter Telescope (LMT) or the Sardinia Radio Telescope (SRT). Note that observations at 50 GHz would be a good time filler when the weather is not very good. \\

\noindent
{\bf References:} Fontani et al. 2024, A\&A, 691A, 180F $\bullet$
Glavin et al. 2025, Nat. Astron., 9, 199 $\bullet$
Jimenez-Serra et al. 2022, A\&A, 663A, 181J $\bullet$
Jim\'enez-Serra et al. 2025, Proceedings of the Kavli-IAU 383 Astrochemistry Symposium,  arXiv:2501.01782 $\bullet$
Kitadai \& Maruyama 2018, Geoscience Frontiers 9, 1117-1153 $\bullet$
Klaassen et al. (2024), Open Res Europe 2024, 4:112 $\bullet$
Li et al. 2017, ApJ, 849, 115L $\bullet$
Li et al. 2020, MNRAS, 492, 556L $\bullet$
Oba et al. 2023, Nat. Comms., 14, 1292 $\bullet$
Ohishi 2019, Astrobiology, Springer Nature Singapore Pte Ltd. $\bullet$
Rivilla et al. 2020, ApJ, 899L, 28R $\bullet$
Rivilla et al. 2021, PNAS, 11801314R $\bullet$ 
Sanz-Novo et al. 2023, ApJ, 954, 3 $\bullet$
Shimonishi et al. 2021, ApJ, 922, 206S $\bullet$
Shimonishi 2024, Proceedings of the Kavli-IAU 383 Astrochemistry
Symposium, arXiv:2411.04451 

\end{document}